\documentstyle[11pt,psfig,aaspp4]{article}
\def\etal{{\it et al. }}
\def\kms{km~s$^{-1}$~}
\def\kmsm{km~s$^{-1}$~Mpc$^{-1}$}
\def\hnot{H$_\circ$~}
\def\vpec{V$_{pec}$~}

\def\W50{W$_{50}$~}
\journalid{337}{}
\articleid{11}{14}

\slugcomment{Submitted to {\it The Astrophysical Journal (Letters)}}

\begin{document}
\title{The Tully--Fisher Relation and \hnot}

\author {Riccardo Giovanelli, Martha P. Haynes}
\affil{Center for Radiophysics and Space Research
and National Astronomy and Ionosphere Center,
Cornell University, Ithaca, NY 14953}

\author {Luiz N. da Costa}
\affil{European Southern Observatory, Karl Schwarzschild Str. 2, D--85748
Garching b. M\"unchen, Germany and Observatorio Nacional, Rio de Janeiro, Brazil}

\author {Wolfram Freudling}
\affil{European Southern Observatory and Space Telescope--European Coordinating Facility, Karl 
Schwarzschild Str. 2, D--85748
Garching b. M\"unchen, Germany}

\author {John J. Salzer}
\affil{Dept. of Astronomy, Wesleyan University, Middletown, CT 06457}

\author {Gary Wegner}
\affil{Dept. of Physics and Astronomy, Dartmouth College, Hanover, 
NH 03755}


\begin{abstract}

The use of the Tully--Fisher (TF) relation for the determination of \hnot
relies on the availability of an adequate template TF relation and of
reliable primary distances. Here we use a TF template relation with the
best available kinematical zero--point, obtained from a sample of 24 
clusters of galaxies extending to $cz \sim 9,000$ \kms, and the most
recent set of Cepheid distances for galaxies fit for TF use. The 
combination of these two ingredients yields H$_\circ = 69\pm5$ \kmsm.
The approach is significantly more accurate than the more common application
with single cluster (e.g. Virgo, Coma) samples.

\end{abstract}


\section {Introduction}

The Tully--Fisher (1977; hereafter TF) relation is a scaling law that correlates 
the luminosity and the rotational velocity of spiral galaxies. As such, it is well 
suited to yield estimates of distance ratios between galaxies or galaxy aggregates. 
Once calibrated by means of a reliable {\it kinematical zero point}, the TF 
relation can be profitably used to measure peculiar velocities \vpec, i.e. 
deviations from smooth Hubble flow. It can also be used as a tool in the 
measurement of the Hubble constant, \hnot (see Jacoby \etal 1992 and refs. 
therein); to that end, however, an additional calibration is needed for the 
absolute magnitude scale. The latter can be obtained from galaxies suitable
for TF use with available primary distances. The number of such objects is rapidly 
growing because of the expansion of the Cepheid horizon made possible by
the Hubble Space Telescope. Optimization of the kinematical calibration has 
received less attention, and we make it our main target in this letter.

Consider the following application.
Suppose fluxes and velocity widths are measured for a sample of galaxies in a 
cluster; an apparent magnitude--velocity width diagram can thus be constructed, 
insofar as all the galaxies in the sample can be assumed to be cluster members. 
Suppose moreover that a set of nearby spiral galaxies have Cepheid distances. 
Matching the TF diagram of the cluster to that of the nearby calibrators yields 
a distance modulus for the cluster; then a measure of its systemic radial velocity 
can yield an estimate of \hnot. The systemic radial velocity is however 

$$cz = H_\circ d + [{\bf V}_{pec}({\bf d}) - {\bf V}_{pec}(0)]\cdot ({\bf d}/d)
\eqno (1)$$

\noindent
where {\bf V}$_{pec}$ is the peculiar velocity vector, {\bf d} is the
vector distance to the galaxy and $d$ its modulus. If the Cosmic
Microwave Background (CMB) radiation dipole moment is interpreted as
a Doppler shift resulting from the motion of the Local Group with respect
to the comoving reference frame, ${\bf V}_{pec}(0)$ can be inferred from
the CMB dipole. Thus, solving for $H_\circ$ still requires knowledge of
the peculiar velocity of the cluster. The Virgo cluster has often been a
target of such application: it is nearby, well studied and  Cepheids
have been discovered in several of its member galaxies. The uncertainty on its 
peculiar velocity is however still large (cf. Freedman \etal 1994) and amounts 
to a sizable fraction of the cluster redshift, so that it percolates
heavily in the derived \hnot error budget. A standard technique to circumvent
this problem is that of bootstrapping the uncertainty to a more distant
cluster, e.g. Coma (Freedman \etal 1994; Tanvir \etal 1995; Yasuda and 
Okamura 1996). Using techniques such as TF that yield reliable
distance ratios, the distance ratio between, say, Virgo and Coma can be
obtained. Since Coma is nearly 6 times farther than Virgo, the ratio between
the uncertain cluster peculiar velocity and its redshift is smaller. The
kinematical calibration mentioned above is thus implicitly obtained by 
assuming that the distant cluster is at rest in the comoving reference frame,
thus introducing a relative error equal to the ratio between the unknown
peculiar velocity and the systemic velocity of the cluster.

The limitations of this approach are numerous. Cluster peculiar velocities
can be large, and an important fraction of the redshift even for Coma. 
Estimates of the amplitude
of cluster peculiar motions still differ. While the recent measurements of
Giovanelli et al. (1996b) yield cluster peculiar motions not exceeding
600 \kms, other sources report motions of significantly larger amplitude 
(Lauer and Postman 1994; see also Moscardini \etal 1996 and Bahcall and Oh 1996).
Moreover, as the distance of the cluster increases, 
so does the amplitude of the incompleteness bias and that of its correction.
Finally, a TF template relation extracted from a single cluster is naturally
restricted to the inclusion of, at most, a few dozen relatively bright
galaxies, limiting the statistical accuracy in the definition of slope and
offset of the TF template relation. Faint objects not only introduce very
large scatter, but also uncertain nonlinearity in TF diagram:
their inclusion in a TF template is thus of negligible or perhaps even 
negative consequence (Giovanelli \etal 1996b).

There is a better way. Rather than using a single cluster in obtaining a TF
template, the kinematical calibration of the TF relation is better achieved 
by using a ``basket of 
clusters''. The quality of the kinematical calibration thus obtained
improves as the sky coverage of the cluster distribution becomes more
isotropic and as the redshift distribution becomes deeper. Such a template
may even be impervious to the possible presence of large--scale bulk flows,
if the sky distribution of clusters is sufficiently isotropic. Moreover,
the global template that can be obtained from such set is richer and has
significantly better defined TF parameters (slope, offset) than those
obtainable from any single cluster template. This approach resembles in
some aspects that followed by Jerjen and Tammann (1993). In Section 2, we 
discuss such a TF template. In section 3, we survey the available literature 
on TF calibrators with  Cepheid distances, and in Section 4 we combine 
template and calibrators to derive an estimate of \hnot.

\section {A TF Template Relation}

In a recent study (Giovanelli \etal 1996a, 1996b; hereafter G96a and G96b),
555 galaxies with TF measurements, located in the fields of 24 clusters, 
were used to produce a global TF template, meant to be used in the estimate 
of peculiar velocities. The size of the cluster galaxy sample allows the 
application of strict membership criteria. The cluster distribution, which
well approximates isotropy and extending to $cz \sim 9000$ \kms, makes 
possible a sensible definition of the kinematical zero point. Figure 1 shows 
the TF diagram from which the template relation is obtained, after a number 
of corrections were applied to the data, including those that take in 
consideration differences among morphological types, cluster sample 
incompleteness bias and cluster motions with respect to the CMB reference
frame. A detailed description of those corrections is given in G96b. A few
points are worth stressing here.

\begin{figure} 
\centerline{\psfig{figure=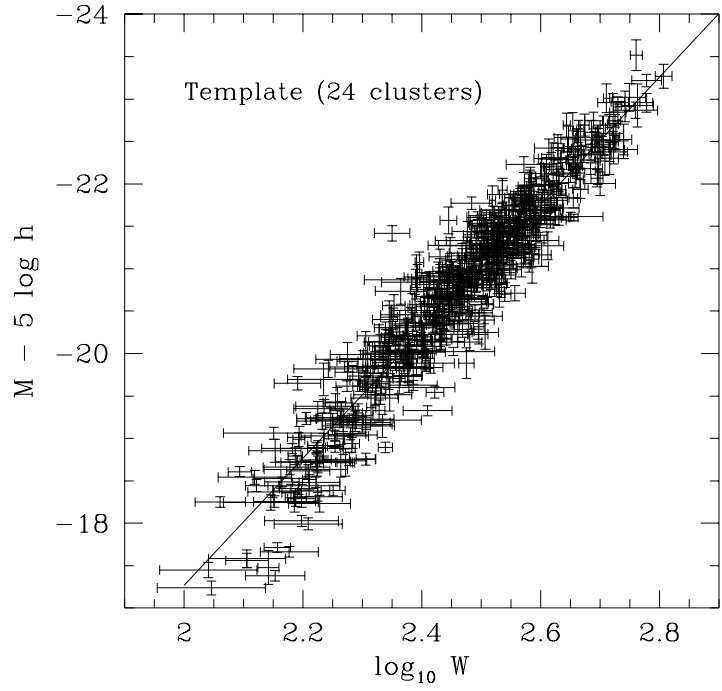,height=6.0in}}
\caption {Template relation based on 555 galaxies in 24 clusters.}
\end{figure}

First, the dispersion about the TF mean relation is not well represented by 
a single figure of scatter. As shown in G96b, the scatter increases steeply 
as the velocity width diminishes. This has a strong effect in the computation 
of bias corrections and therefore on the vertical offset of the template. 
Second, the mean amplitude of the scatter is about
0.35 mag. Since the amplitude of the bias corrections scales with the 
scatter amplitude, the corrections we have estimated are smaller than those
advocated by Sandage (1994), who finds a significantly larger mean scatter
in the TF relation. Third, given the characteristics of the scatter, which
are different from those assumed by Schechter (1980) and adopted by others, 
i.e. their dependence on location in the TF plane, on measurement and correction 
errors as well as on other cosmic sources, the idea that an inverse TF fit 
(i.e. one where $\log W$ is used as the ``dependent'' variable) is a bias--free  
tool is not well justified. The bias corrections applied 
to the data in Figure 1 were estimated within the framework of a  
bivariate TF fit, where magnitude was used as the dependent variable, and 
errors in both coordinates were taken into account. The adopted  
bivariate fit to the data, as plotted in the figure, is

$$M_I + 5 \log h = -21.00\pm0.02 - 7.68\pm0.13 (\log W - 2.5) \eqno (2)$$

\noindent
where $h=H_\circ/100$ in the usual units. Eqn. (2) 
allows the prediction of $M_I + 5 \log h$ from a galaxy's velocity width; its 
combination with the measured flux then yields an estimate of the distance, 
scaled by $h$, which can be replaced in eqn. (1) to obtain the peculiar velocity.
The kinematical calibration relies on how well the weighted mean of the
peculiar velocities of the set of clusters approaches a null value. Note
that even in the presence of a linear bulk flow, an isotropically distributed 
cluster set will yield a null mean velocity.

In addition to the statistical errors indicated in eqn. (2), which arise from the
TF scatter characteristics and the size of the sample, other sources of uncertainty
affect the quality of the TF template, in the form of systematic errors. The 
first of those additional contributions arises from the uncertainty in the
cluster bias corrections, which depend on the assumed shape of the galaxy
luminosity function. In G96b we estimate that contribution to the uncertainty of the
template zero point to be about 0.03 mag. The second, and more important term
depends on how good the determination of the kinematical zero point is for the
cluster sample. If in the cluster peculiar velocity distribution function 
peculiar velocities in excess of $\sim 600$ \kms are rare, as suggested by
the results in G96b and illustrated in Bahcall and Oh (1996), the r.m.s.
departure from null velocity expected for an isotropic cluster average will translate
in a TF template zero point uncertainty of about 0.04 mag, as discussed in
G96b. If, on the other hand, a broader range of peculiar velocities is
allowed (as shown, e.g. in Moscardini \etal 1996), its effect on the 
uncertainty of the TF zero point may be as large as 0.06 mag. The combination
of statistical, luminosity function--related and kinematical uncertainties
on the TF offset can thus amount to as much as 0.05 to 0.07 mag. This is 
significantly lower than, for example a 10\% ($\sim 0.2$ mag) uncertainty 
due to the unknown peculiar velocity of a single cluster.

\section {TF Galaxies with Cepheid Distances}

The number of spiral galaxies with distance determination via the Cepheid
period--luminosity relation has grown rapidly in the last couple of years, 
precipitating a rush of estimates of \hnot of increasing accuracy (see
papers in Livio \etal 1996, especially those by Freedman, Mould and Tammann).
As a result, the number of ``Cepheid calibrators'' of the TF relation
has also increased substantially. The TF template relation discussed in 
section 2 refers to fluxes gathered in the I band. In Table 1, we include
galaxies with measured I band magnitudes, valid estimates of the velocity 
width and Cepheid distances known to us. The I band magnitudes (col. 5) are 
mostly those measured by Pierce and Tully (1992) and Pierce (1996), except 
for NGC 1365, for which we use the average magnitude between those published
by Mathewson \etal (1992) and by Bureau \etal (1996). Velocity widths (col. 11)
have been garnered from many sources in the public domain, and are discussed
case by case below. The corrections applied to the
observed values, to account for disk inclination, turbulence contributions, 
etc., are those described in G96a. The morphological type correction $\beta_{typ}$,
listed in col. 8, is an term to be subtracted from the magnitude, that takes into 
consideration mean offsets of galaxies of different types from those of type 
Sbc/Sc, as discussed in G96b. The adopted distance moduli are listed in col. 9. 
Galactic and internal extinctions (cols. 6 and 7) are estimated using the 
procedures described in G96a. NGC and Messier numbers are given in cols. 1--2, 
while RC3 type and inclination are listed in cols. 3 and 4.


Many of the objects listed in Table 1 are not ideal for TF use. M100, M101
and N4496 are uncomfortably close to face--on, requiring large inclination
corrections to the widths. M31, M33 and M81 exhibit severe perturbations in
the velocity field. NGC 2366 and NGC 3109 are dwarf irregular systems, very
ill--suited for use with a template that is principally constructed using
luminous spirals; for dwarf systems, moreover, it is difficult to estimate
accurately the actual contribution to the width of rotational motions, and
their scatter about the TF relation is huge (Hoffman and Salpeter 1996). 
NGC 4496 is not only nearly face--on, but a second galaxy is seen superimposed
on its disk, making the extraction of photometric parameters from any image
highly uncertain.
We note one further caution. Our estimates of TF parameters in 
Table 1 are based on a somewhat subjective synthesis of a large amount of
heterogeneous material, especially concerning the velocity widths, sometimes 
involving the measurement of spectra on paper copies of published data 
figures. Our assignment of error bars to the data is thus reflective of 
this unorthodox method of parameter derivation, rather than of the original 
accuracy of the data, and it is likely to underestimate somewhat the
amplitude of the uncertainty.

\begin{figure} 
\centerline{\psfig{figure=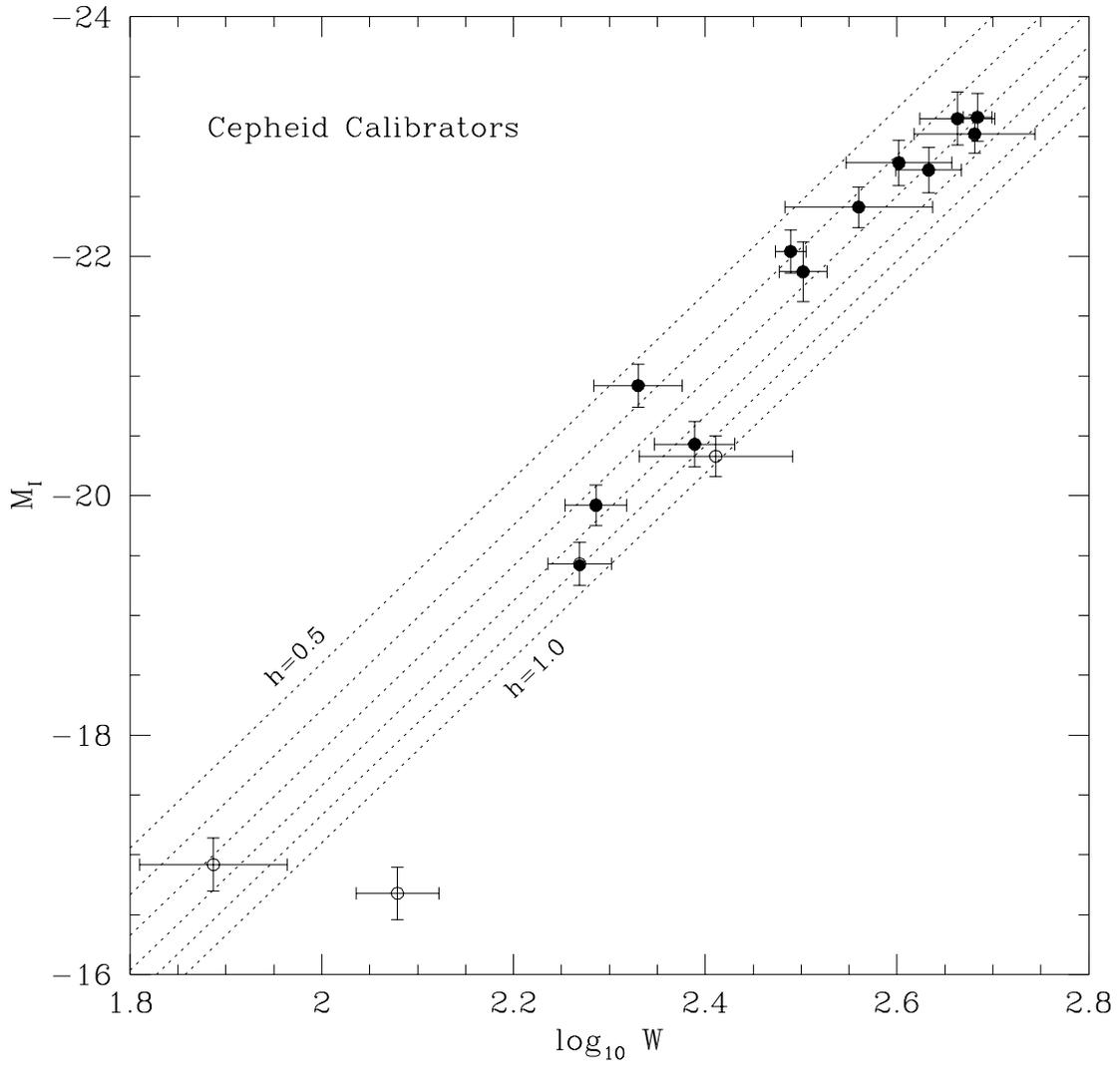,height=6.0in}}
\caption {TF calibrators with Cepheid distances, superimposed on
a grid of TF template relations plotted for different values of $h$.
The unfilled symbols represent galaxies unfit for TF work, as 
explained in the text.}
\end{figure}

\section {The Value of \hnot}

In Figure 2, the data of all galaxies listed in Table 1 are plotted over
a grid of renditions of the TF template relation given in eqn. (2), each
for a different value of $h$ between 0.5 and 1.0. According
to the comments in Section 3, three of the objects: NGC 4496, NGC 2366 and
NGC 3109 are most ill--suited for TF use, and are plotted by unfilled
symbols in Figure 2. We do not use them as calibrators, restricting our set to
a total of 12 galaxies. Keeping the slope of the TF template fixed, we
compute the value of $h$ that yields $\chi^2$ minimization of residuals
for the set of calibrators. The best fit 
yields $h = 0.69\pm0.02$. The formal error of $\sim 0.06$ mag does not take 
in consideration the systematic uncertainty on the TF relation zero point
discussed in section 2, nor possible systematic uncertainties on the 
period--luminosity relation of Cepheid distances. Allowing for such an
``external'' uncertainty of about 0.1 mag, the combination in quadrature
of all discussed sources of error yields an uncertainty of $\sim 0.16$ mag, or

$$H_\circ = 69\pm5 ~~{\rm km s}^{-1}~{\rm Mpc}^{-1}~~\eqno (3)$$

\noindent
This result is more robust than those derived purely from distances
to galaxies in Virgo, or M96, or from applications that rely on bootstraps 
of the distances of those aggregates to that of Coma.

\acknowledgements

It is a pleasure to acknowledge stimulating conversations with Mort Roberts, 
on the rotational velocity of M31, and with Michael Pierce, who also unselfishly 
contributed results ahead of publication.
The results presented in this paper are in part based on observations carried out at
the National Astronomy and Ionosphere Center (NAIC), the National Radio Astronomy
Observatory (NRAO), the Kitt Peak National Observatory (KPNO), the Cerro Tololo 
Interamerican Observatory (CTIO), the Palomar Observatory (PO), the Observatory of 
Paris at Nan\c cay, the Michigan--Dartmouth--MIT Observatory (MDM) and the European
Southern Observatory (ESO). NAIC, NRAO, KPNO and CTIO are respectively operated by 
Cornell University, Associated Universities, inc., and Associated Universities for 
Research in Astronomy, all under cooperative agreements with the National Science 
Foundation. Access to the 5m telescope at PO is guaranteed under an agreement 
between Cornell University and the California Institute of Technology. 
This research was supported by NSF grants AST94--20505 to RG, AST90-14850 and AST90-23450 to MH and AST93--47714 to GW.

\end{document}